\documentclass[10pt,a4paper]{article}
\usepackage[utf8]{inputenc}
\usepackage{amsmath,amssymb}
\usepackage{graphicx}

\usepackage[margin=2.5cm]{geometry}

\usepackage{color}

\usepackage{IEEEtrantools}
\usepackage{bm}
\usepackage{hyperref}

\usepackage{mathtools}

\usepackage{booktabs}

\usepackage{siunitx}

\usepackage[acronym, style=alttree, toc=true, shortcuts, xindy, nomain, nonumberlist]{glossaries}

\usepackage{hyphenat}

\newcommand{\bs}[1]{\boldsymbol{#1}}

\newcommand{\lr}[1]{\left(#1\right)} 

\DeclarePairedDelimiterXPP\onenorm[1]{}\lVert\rVert{_1}{\ifblank{#1}{\:\cdot\:}{#1}} 
\DeclarePairedDelimiterXPP\twonorm[1]{}\lVert\rVert{_2}{\ifblank{#1}{\:\cdot\:}{#1}} 

\newglossary{symbols}{sym}{sbl}{List of Symbols}
\newglossary{notation}{not}{nt}{Notation}

\glsaddkey{body}{\glsentrytext{\glslabel}}{\glsentrybody}{\GLsentrybody}{\glsbody}{\Glsbody}{\GLSbody}

\newacronym{ocp}{OCP}{optimal control problem}
\newacronym{MPC}{MPC}{Model Predictive Control}
\newacronym{cvpm}{CVPM-MPC}{CVPM-MPC}
\newacronym{CVPM}{CVPM}{constraint violation probability minimization}
\newacronym{RMPC}{RMPC}{Robust Model Predictive Control}
\newacronym{SMPC}{SMPC}{Stochastic Model Predictive Control}
\newacronym{SCMPC}{SCMPC}{Scenario Stochastic Model Predictive Control}
\newacronym{pdf}{pdf}{probability density function}
\newacronym{cdf}{cdf}{cumulative distribution function}

\newglossaryentry{input}{type=symbols,
	sort={u1},
	name={\ensuremath{\bs{u}_k}},
	plural={\ensuremath{\bs{u}}},
	description={Input vector with the dimension \gls{dimu} at time $k$},
}
\newglossaryentry{state}{type=symbols,
	sort={x1},
	name={\ensuremath{\bs{x}_k}},
	plural={\ensuremath{\bs{x}}},
	description={State vector with the dimension \gls{dimx} at time $k$}
}
\newglossaryentry{rp}{type=symbols,
	sort={p},
	name={\ensuremath{\beta}},
	plural={\ensuremath{\beta}},
	description={Risk parameter}
}
\newglossaryentry{rp_brackets}{type=symbols,
	sort={p},
	name={\ensuremath{\beta}},
	plural={\ensuremath{\beta}},
	description={Risk parameter}
}

\newglossaryentry{ominus}{type=notation,
	sort={pontriangin},
	name={\ensuremath{\ominus}},
	description={Pontryagin difference}
}	

\newglossaryentry{oplus}{type=notation,
	sort={Minkowski},
	name={\ensuremath{\oplus}},
	description={Minkowski sum}
}

\title{(Stochastic) Model Predictive Control - a Simulation Example}
\author{Tim Brüdigam\\ \small Technical University of Munich, 80333 Munich, Germany\\ \tt\small tim.bruedigam@tum.de}
\date{19.04.2021}

\begin{document}

\maketitle

\begin{abstract}
    This brief introduction to Model Predictive Control specifically addresses stochastic Model Predictive Control, where probabilistic constraints are considered. A simple linear system subject to uncertainty serves as an example. The Matlab code for this stochastic Model Predictive Control example is available online.
\end{abstract}

\tableofcontents

\newpage

\section{Introduction}

In the following, we provide details on an (S)MPC simulation example. The Matlab code for this simulation example is available at \url{https://github.com/tim283/smpc_example}. The main purpose of this document is to introduce the idea of considering uncertainty in constraints within \gls{MPC}. This document is not necessarily a step-by-step manual, nor does it explain code in detail. 

In Section~\ref{sec:simu}, we first introduce the deterministic system, constraints, and MPC optimal control problem. We then introduce uncertainty into the system dynamics and provide a brief overview of how to handle constraints subject to uncertainty, also called chance constraints. 

Section~\ref{sec:SMPC} provides a more elaborate derivation of the chance constraint reformulation, both for normally distributed uncertainties and general probability distributions.

\section{Simulation Example}
\label{sec:simu}

We consider the system example described in \cite{LorenzenEtalAllgoewer2017} to illustrate the concept of MPC. Based on the MPC toolbox of \cite{GruenePannek2017}, the introduced system may then be controlled.

While this example system is considered here without any physical context, the system represents a linearization of a Buck-Boost DC–DC converter \cite{LazarEtalvandenBosch2008, CannonEtalCheng2011}.

\subsection{System and Constraints}

We first consider the system
\begin{IEEEeqnarray}{c}
\bm{x}_{k+1} = \bm{A}\bm{x}_k + \bm{B}u_k
\end{IEEEeqnarray}
with states $\bm{x}_k = [ x_{1,k},~x_{2,k}]^\top$ and input $u_k$. Here, we choose
\begin{IEEEeqnarray}{c}
\bm{A} = \begin{bmatrix}
1& 0.0075\\ -0.143 & 0.996
\end{bmatrix},~~
\bm{B} = \begin{bmatrix}
4.798\\ 0.115
\end{bmatrix}
\end{IEEEeqnarray}
for the system and input matrix. In addition, we consider input constraints
\begin{IEEEeqnarray}{c}
-0.2 \leq u_k \leq 0.2
\end{IEEEeqnarray}
and a state constraint limiting the state $x_1$ by
\begin{IEEEeqnarray}{c}
x_{1,k} \leq x_{1,\text{lim}} \label{eq:stateconstraint}
\end{IEEEeqnarray}
where $x_{1,\text{lim}} = 2.8$ is the limit. 

\subsection{Model Predictive Control}

The MPC optimal control problem, solved at each time step, subject to only input constraints is given by
\begin{IEEEeqnarray}{rll}
\IEEEyesnumber \label{eq:mpc_problem_stateC}
 \min_{\bm{U}} &\sum_{k=0}^{N-1} \bm{x}_k^\top \bm{Q} \bm{x}_k + u_k^\top \bm{R} u_k & \IEEEyessubnumber\\
\textnormal{s.t. }& \bm{x}_{k+1} = \bm{A}\bm{x}_k + \bm{B}u_k, & ~k \in \{ 0, ..., N-1 \} \IEEEyessubnumber \\
& |u_k| \leq 0.2, & ~k \in \{ 0, ..., N-1 \} \IEEEyessubnumber
\end{IEEEeqnarray}
with $\bm{U} = [u_0, ... u_{N-1}]$, horizon $N=11$, and sampling time $\Delta t = 0.1$, where the weighting matrices are
\begin{IEEEeqnarray}{c}
\bm{Q} = \begin{bmatrix}
1& 0\\ 0 & 10
\end{bmatrix},~~
\bm{R} = 1 .
\end{IEEEeqnarray}

If no state constraints are present, starting at the initial state $\bm{x}_0 = [2.5,~4.8]^\top$ results in a curved motion towards the origin, where $x_1$ first increases (beyond the value $2.8$). By introducing the state constraint \eqref{eq:stateconstraint}, the MPC optimal control problem is extended to
\begin{IEEEeqnarray}{rll}
\IEEEyesnumber
 \min_{\bm{U}} &\sum_{k=0}^{N-1} \bm{x}_k^\top \bm{Q} \bm{x}_k + u_k^\top \bm{R} u_k & \IEEEyessubnumber\\
\textnormal{s.t. }& \bm{x}_{k+1} = \bm{A}\bm{x}_k + \bm{B}u_k, & ~k \in \{ 0, ..., N-1 \} \IEEEyessubnumber \\
& |u_k| \leq 0.2, & ~k \in \{ 0, ..., N-1 \} \IEEEyessubnumber \\
& x_{1,k} \leq 2.8, & ~k \in  \{1, ..., N \} \IEEEyessubnumber 
\end{IEEEeqnarray}
causing $x_1$ to increase until the value $2.8$ is reached. Then, the states move towards the origin along the constraint.

\subsection{Stochastic MPC for Systems with Uncertainty}

Now, uncertainty is introduced into the system, yielding
\begin{IEEEeqnarray}{c}
\bm{x}_{k+1} = \bm{A}\bm{x}_k + \bm{B}u_k + \bm{D} \bm{w}_k \label{eq:sys_unc}
\end{IEEEeqnarray}
with 
\begin{IEEEeqnarray}{c}
\bm{D} = \begin{bmatrix}
1& 0\\ 0 & 1
\end{bmatrix}.
\end{IEEEeqnarray}
and the Gaussian uncertainty $\bm{w}_k \sim \mathcal{N} (\bm{0}, \bm{\Sigma}_w)$ with covariance matrix 
\begin{IEEEeqnarray}{c}
\bm{\Sigma}_w = \begin{bmatrix}
0.08& 0\\ 0 & 0.08
\end{bmatrix}.
\end{IEEEeqnarray}

Assuming we now used~\eqref{eq:mpc_problem_stateC} to control the actual system~\eqref{eq:sys_unc}, the MPC would not account for uncertainty and the state constraint may be violated, as the prediction model does not account for the actual uncertainty. In order to cope with this, an SMPC approach is employed, accounting for the uncertainty $\bm{w}_k$. Therefore, the state constraint is transformed into a chance constraint
\begin{IEEEeqnarray}{c}
\mathrm{Pr}(x_{1,k} \leq 2.8) \geq \gls{rp} \label{eq:cc}
\end{IEEEeqnarray}
where $\gls{rp}$ is a risk parameter. The constraint \eqref{eq:stateconstraint} is not required to hold always, but only up to a level specified by the predefined risk parameter. The higher the risk parameter is chosen, the lower the risk allowed. Sometimes, the risk parameter is defined as $\tilde{\gls{rp}} = 1 - \gls{rp}$, where \gls{rp} in \eqref{eq:cc} is replaced with $1-\tilde{\gls{rp}}$. It then holds that increasing the risk parameter $\tilde{\gls{rp}}$ also increases risk.

This chance constraint is required to be reformulated in order to be used within the optimal control problem, yielding
\begin{IEEEeqnarray}{rl}
\IEEEyesnumber
& x_{1,k} \leq 2.8 - \gamma \IEEEyessubnumber \\
& \gamma = \sqrt{2 [1,0]^\top \bm{\Sigma}^\text{e}_k [1,~0]}~ \mathrm{erf}^{-1}(2\gls{rp_brackets} -1) \IEEEyessubnumber
\end{IEEEeqnarray}
which is an approximation of \eqref{eq:cc} \cite{FarinaGiulioniScattolini2016}. In other words, the state constraint is tightened by the tightening parameter $\gamma$, which itself depends on the risk parameter $\gls{rp}$ and the error covariance matrix $\bm{\Sigma}^\text{e}_k$ that is derived in the following. The vector $[1,0]$ is required as only the error affecting $x_1$ is important. A more extensive derivation of the chance constraint reformulation is provided in Section~\ref{sec:SMPC}. In short, the term $\sqrt{[1,0]^\top \bm{\Sigma}^\text{e}_k [1,~0]}$ considers the uncertainty (and the resulting error) variance and $\sqrt{2 }~ \mathrm{erf}^{-1}(2\gls{rp_brackets} -1)$ follows from the uncertainty distribution (here the normal distribution) and the risk parameter.

The state of the system dynamics \eqref{eq:sys_unc} may be split into a deterministic and a probabilistic part
\begin{IEEEeqnarray}{c}
\bm{x}_k = \bm{z}_k + \bm{e}_k
\end{IEEEeqnarray}
and the input is adapted to
\begin{IEEEeqnarray}{c}
u_k = -\bm{K}\bm{x}_k + v_k
\end{IEEEeqnarray}
where $\bm{K}$ is a stabilizing state feedback matrix obtained by an LQR approach. Here, $\bm{K}\bm{x}_k$ controls the deterministic part of the system dynamics, while the new decision variable $v_k$ accounts for uncertainty.

MPC requires propagating the error for the prediction, yielding the error covariance matrix
\begin{IEEEeqnarray}{c}
\bm{\Sigma}^\text{e}_{k+1} = \left(\bm{A} - \bm{B} \bm{K}  \right) \bm{\Sigma}^\text{e}_k \left(\bm{A} - \bm{B} \bm{K}  \right)^\top + \bm{D} \bm{\Sigma}_w {\bm{D}}^\top
\end{IEEEeqnarray}
for every prediction step with 
\begin{IEEEeqnarray}{c}
\bm{\Sigma}^\text{e}_0 = \mathrm{diag}(0,0).
\end{IEEEeqnarray}
Therefore, the error covariance matrix depends on the previous error propagated through the system, as well as the covariance of the uncertainty additionally added at each step. Note the difference between the uncertainty covariance matrix $\bm{\Sigma}_w$ and the error covariance matrix $\bm{\Sigma}^\text{e}_{k}$, which is computed for every prediction step $k$, based on the prediction model and on $\bm{\Sigma}_w$.

The full SMPC optimal control problem is
\begin{IEEEeqnarray}{rll}
\IEEEyesnumber \label{eq:SMPC_ocp}
 \min_{\bm{V}} &\sum_{k=0}^{N-1} \bm{x}_k^\top \bm{Q} \bm{x}_k + (v_k - \bm{K}\bm{x}_k)^\top \bm{R} (v_k - \bm{K}\bm{x}_k) & \IEEEyessubnumber\\
\textnormal{s.t. }& \bm{x}_{k+1} = \bm{A}\bm{x}_k + \bm{B} (v_k - \bm{K}\bm{x}_k), & k \in \{ 0, ..., N-1 \} \IEEEyessubnumber \\
& |v_k - \bm{K}\bm{x}_k| \leq 0.2, & k \in \{ 0, ..., N-1 \} \IEEEyessubnumber \\
& x_{1,k} \leq 2.8 - \gamma_k, & k \in  \{1, ..., N \} \IEEEyessubnumber \\
& \gamma_k = \sqrt{2 [1,0] \bm{\Sigma}^\text{e}_k [1,~0]^\top}~ \mathrm{erf}^{-1}(2\gls{rp_brackets} -1), & k \in  \{1, ..., N \} \IEEEyessubnumber
\end{IEEEeqnarray}
where $\bm{V} = [v_0, ... v_{N-1}]$ replaces the decision variables $\bm{U}$. 

If \eqref{eq:SMPC_ocp} is applied to control the actual system \eqref{eq:sys_unc}, the states do not exactly reach the state constraint, but a margin is established, providing space to account for uncertainty (the constraint is tightened). Increasing the risk parameter $\gls{rp}$ increases the required distance towards the state constraint, again resulting in less constraint violation.

\begin{IEEEeqnarray}{rll}
\IEEEyesnumber
 \min_{\bm{U}} &\sum_{k=0}^{N-1} \bm{x}_k^\top \bm{Q} \bm{x}_k + u_k^\top \bm{R} u_k & \IEEEyessubnumber\\
\textnormal{s.t. }& \bm{x}_{k+1} = \bm{A}\bm{x}_k + \bm{B}u_k, & ~k \in \{ 0, ..., N-1 \} \IEEEyessubnumber \\
& |u_k| \leq 0.2, & ~k \in \{ 0, ..., N-1 \} \IEEEyessubnumber \\
& x_{1,k} \leq 2.8, & ~k \in  \{1, ..., N \} \IEEEyessubnumber 
\end{IEEEeqnarray}

\section{Details on SMPC with Chance Constraints}
\label{sec:SMPC}

In the following, a brief derivation of chance constraints in SMPC is provided. First, the system, constraint, and chance constraint setup are introduced.

We consider a linear system with additive uncertainty $\bm{w}_k $
, i.e.,
\begin{IEEEeqnarray}{c}
\bm{x}_{k+1} = \bm{A} \bm{x}_k + \bm{B} \bm{u}_k + \bm{D} \bm{w}_k .
\end{IEEEeqnarray}

The system state is split into a deterministic state $\bm{z}_k$ and a probabilistic error $\bm{e}_k$. The input is also split into two parts, a feedback law stabilizing the deterministic system, as well as a new input $\bm{v}_k$ accounting for uncertainty. These considerations yield
\begin{IEEEeqnarray}{rl}
\IEEEyesnumber
\bm{x}_k &= \bm{z}_k + \bm{e}_k \IEEEyessubnumber\\
\bm{u}_k &= \bm{K}\bm{x}_k + \bm{v}_k \IEEEyessubnumber
\end{IEEEeqnarray}

Splitting the system state results in the new system models
\begin{IEEEeqnarray}{rl}
\IEEEyesnumber
\bm{z}_{k+1} &= \bm{\Phi} \bm{z}_k + \bm{B} \bm{v}_k \IEEEyessubnumber \\
\bm{e}_{k+1} &= \bm{\Phi} \bm{e}_k + \bm{D} \bm{w}_k \IEEEyessubnumber
\end{IEEEeqnarray}
with $\bm{\Phi} = \bm{A} + \bm{B}\bm{K}$.

We consider the linear constraint
\begin{IEEEeqnarray}{c}
\bm{g}^\top_k \bm{x}_k \leq h_k. \label{eq:linconstraint}
\end{IEEEeqnarray}
Comparing \eqref{eq:linconstraint} to the simulation example in Section~\ref{sec:simu}, we obtain $\bm{g}_k^\top = [1, 0]^\top$ and $h_k = 2.8$. 

As the system state is subject to (unbounded) uncertainty, it is not possible to guarantee constraint satisfaction in all cases. Therefore, we introduce a probabilistic constraint (chance constraint), which bounds the constraint violation probability depending on a risk parameter \gls{rp}, yielding
\begin{IEEEeqnarray}{c}
\mathrm{Pr} \left(\bm{g}^\top_k \bm{x}_k \leq h_k \right) \geq \gls{rp} .
\end{IEEEeqnarray}

With the deterministic and error states, we obtain
\begin{IEEEeqnarray}{c}
\mathrm{Pr} \left(\bm{g}^\top_k \bm{z}_k + \bm{g}^\top_k \bm{e}_k \leq h_k \right) \geq \gls{rp}  . \label{eq:cc_split}
\end{IEEEeqnarray}

The chance constraint \eqref{eq:cc_split} is not a deterministic expression. It is necessary to reformulate the chance constraint, such that a tractable expression is obtained, which is then used to solve the optimization problem. For zero uncertainty, the deterministic part of the state must satisfy the state constraint. However, if uncertainty is present, the constraint must be tightened by a tightening parameter $\gamma_k$. This tightening parameter is determined depending on the uncertainty and the risk parameter \gls{rp}. We therefore reformulate the chance constraint \eqref{eq:cc_split} into
\begin{IEEEeqnarray}{rl}
\label{eq:cc_formulation}
\IEEEyesnumber
\bm{g}^\top_k \bm{z}_k &\leq h_k - \gamma_k \IEEEyessubnumber \\
\mathrm{Pr} \left(\bm{g}^\top_k \bm{e}_k \leq \gamma_k \right) &= \gls{rp} \IEEEyessubnumber \label{eq:cc_error}
\end{IEEEeqnarray}
where \eqref{eq:cc_error} ensures that the tightening parameter is chosen in such a way, that the uncertainty only causes constraint violations as allowed by the risk parameter.

It is to note that \eqref{eq:cc_formulation} is still not a deterministic expression. In the following, we derive how \eqref{eq:cc_error} is reformulated into a deterministic approximation, based on the underlying uncertainty distribution. 

\subsection{Normally Distributed Uncertainty}

First, we analyze the chance constraint reformulation for an uncertainty $\bm{w}_k$ subject to a normal distributions with zero mean, i.e, $\bm{w}_k \sim \mathcal{N}\left(0,\bm{\Sigma}\right)$. Similar reformulations are also used in \cite{BlackmoreOnoWilliams2011, OnoWilliams2008, FarinaEtalScattolini2015, CarvalhoEtalBorrelli2014, BruedigamEtalWollherr2018b}.

Given the normal distribution with zero mean, the error is also normally distributed according to
\begin{IEEEeqnarray}{c}
\bm{e}_k \sim \mathcal{N}\left(0,\bm{\Sigma}^e_k\right)
\end{IEEEeqnarray}
with $\bm{\Sigma}^e_0 = 0\bm{I}$ and
\begin{IEEEeqnarray}{c}
\bm{\Sigma}_{k+1}^e = \bm{\Phi} \bm{\Sigma}_k^e \bm{\Phi}^\top + \bm{D} \bm{\Sigma}_w \bm{D}^\top .
\end{IEEEeqnarray}

It follows that the constraint state expression is also normally distributed according to
\begin{IEEEeqnarray}{c}
\bm{g}^\top \bm{e}_k \sim \mathcal{N}\left(0,\bm{g}^\top \bm{\Sigma}^e_k \bm{g}\right)
\end{IEEEeqnarray}
where we assume that $\bm{g}^\top \bm{e}_k$ is scalar and abbreviate the variance with $\sigma^2 = \bm{g}^\top \bm{\Sigma}^e_k \bm{g}$

\subsubsection{Chance constraint reformulation}

Based on the \gls{cdf} of a normal distribution
\begin{IEEEeqnarray}{c}
\mathrm{Pr} \left(\bm{g}^\top_k \bm{e}_k \leq \gamma_k \right) = \frac{1}{2} \left[ 1 + \mathrm{erf} \left( \frac{\gamma_k}{\sigma \sqrt{2}} \right) \right] = \gls{rp}
\end{IEEEeqnarray}
it is possible to find a deterministic expression for the tightening parameter $\gamma_k$ with the following reformulation:
\begin{IEEEeqnarray}{rl}
\IEEEyesnumber
\frac{1}{2} \left[ 1 + \mathrm{erf} \left( \frac{\gamma_k}{\sigma \sqrt{2}} \right) \right] &= \gls{rp} \IEEEyessubnumber \\
\mathrm{erf} \left( \frac{\gamma_k}{\sigma \sqrt{2}} \right) &= 2\gls{rp_brackets} - 1 \IEEEyessubnumber \\
\frac{\gamma_k}{\sigma \sqrt{2}} &= \mathrm{erf}^{-1}(2\gls{rp_brackets} - 1) \IEEEyessubnumber \\
\gamma_k & = \sqrt{2} \sigma~ \mathrm{erf}^{-1}(2\gls{rp_brackets} - 1) \IEEEyessubnumber 
\end{IEEEeqnarray}
Details on the \gls{cdf} of a normal distribution are found in Appendix~\ref{sec:norm_dist}.

Inserting the variance $\sigma^2 = \bm{g}^\top \bm{\Sigma}^e_k \bm{g}$, we obtain the tightening parameter
\begin{IEEEeqnarray}{rl}
\gamma_k  &= \sqrt{2} \sqrt{\sigma^2}~ \mathrm{erf}^{-1}(2\gls{rp_brackets} - 1) \IEEEyessubnumber \\
 &= \sqrt{2} \sqrt{\bm{g}^\top \bm{\Sigma}^e_k \bm{g}}~ \mathrm{erf}^{-1}(2\gls{rp_brackets} - 1) \IEEEyessubnumber \\
  &= \sqrt{2\bm{g}^\top \bm{\Sigma}^e_k \bm{g}}~ \mathrm{erf}^{-1}(2\gls{rp_brackets} - 1) . \IEEEyessubnumber \label{eq:gamma_normdist}
\end{IEEEeqnarray}

\subsubsection{Discussion}

In general, the risk parameter is bounded by $0.5 \leq \gls{rp} < 1$. If the state is exactly on the constraint, given the normal distribution, the probability of violating the constraint in the next step without tightening the constraint is exactly $50\%$. Therefore, no constraint tightening corresponds to a risk parameter of $\gls{rp} = 0.5$. A risk parameter $\gls{rp} = 1$ would guarantee constraint satisfaction. Given the unbounded uncertainty due to the normal distribution, it would follow that $\gamma_k = \infty$ (as $\mathrm{erf}^{-1}(1) = \infty$). Obviously, this is not practical. While it is mathematically possible to choose $\gls{rp} < 0.5$, this does not make sense, as this equals loosening the original hard constraint (constraint loosening instead of constraint tightening).

Furthermore, it is to note that the resulting constraint satisfaction may by larger than specified by the risk parameter, depending on the system and constraints. At first glance, this is surprising, as a normal distribution allows finding an exact reformulation of the chance constraint. However, the tightening parameter $\gamma_k$ increases with increasing prediction steps. To account for this, the realized tightening (due to predicting multiple MPC steps) may be larger than the tightening $\gamma_1$ required to hold for the next step. The chance constraint satisfaction considers only individual prediction steps, not the probability of violating the constraint over an entire trajectory.

\subsection{General Probability Distributions}

A chance constraint reformulation is also possible for general probability distributions. Here, we consider univariate distributions with zero mean and variance $\sigma_w^2$. While a normal distribution allows an exact reformulation, chance constraints for general distributions may only be approximated. An overview is found in \cite{FarinaGiulioniScattolini2016}.

Based on the Cantelli's inequality, see Appendix~\ref{sec:chebyshev}, it is possible to determine a bound on the \gls{cdf}, yielding
\begin{IEEEeqnarray}{c}
\mathrm{Pr} \left(e_k < \gamma_k \right) \geq 1 - \frac{\sigma^2}{\gamma_k^2 + \sigma^2} = \gls{rp}
\end{IEEEeqnarray}
where the risk parameter represents the required bound and $\sigma$ represents the variance of the error, similar to $\bm{\Sigma^e}$ before. 

Reformulating results in
\begin{IEEEeqnarray}{rl}
\IEEEyesnumber
 1 - \frac{\sigma^2}{\gamma_k^2 + \sigma^2} &= \gls{rp} \IEEEyessubnumber \\
\frac{\sigma^2}{\gamma_k^2 + \sigma^2} &= 1 - \gls{rp} \IEEEyessubnumber \\
\sigma^2 &= (1-\gls{rp}) (\gamma_k^2 + \sigma^2) \IEEEyessubnumber \\
\sigma^2 &= \gamma_k^2 (1-\gls{rp}) + \sigma^2 (1-\gls{rp}) \IEEEyessubnumber  \\
\sigma^2 + \sigma^2 (\gls{rp}-1) &= \gamma_k^2 (1-\gls{rp})  \IEEEyessubnumber  \\
\sigma^2(1+\gls{rp}-1) &= \gamma_k^2 (1-\gls{rp})  \IEEEyessubnumber \\
\gamma_k^2 &= \sigma^2 \frac{\gls{rp}}{1-\gls{rp}} \IEEEyessubnumber \\
\gamma_k &= \sigma \sqrt{\frac{\gls{rp}}{1-\gls{rp}}} \IEEEyessubnumber \label{eq:gamma_gen}
\end{IEEEeqnarray}
representing a constraint tightening parameter for general probability distributions. We can rewrite this expression as
\begin{IEEEeqnarray}{c}
\gamma_k = \sigma Q(\gls{rp})
\end{IEEEeqnarray}
where the quantile function $Q(\gls{rp})$ may be considered as the inverse \gls{cdf}.

Unlike the reformulation for the norm constraint, the risk parameter here may take values $0 \leq \gls{rp} < 0$.

\begin{figure}[h]
\centering
\includegraphics[scale=1]{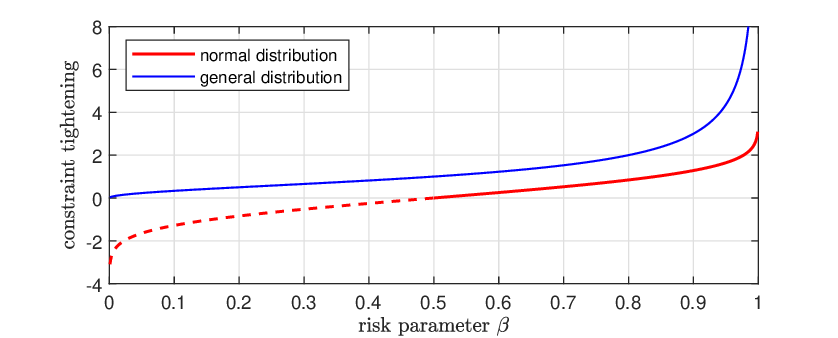}
\caption{Constraint tightening comparison}
\label{fig:tightening_comp}
\end{figure}

A comparison between constraint tightening based on \eqref{eq:gamma_normdist}, i.e, $\sqrt{2}\mathrm{erf}^{-1}(2\gls{rp_brackets} - 1)$, and \eqref{eq:gamma_gen}, i.e., $\sqrt{\gls{rp} / (1-\gls{rp})}$, is shown in Figure~\ref{fig:tightening_comp}. Note that the constraint tightening in Figure~\ref{fig:tightening_comp} is not $\gamma_k$, as variance is not considered. It is obvious that the constraint tightening for general probability distributions according to \eqref{eq:gamma_gen} is more conservative.

\subsection{Normal Distributions with Non-zero Mean}

For non-zero mean probability distributions, the chance constraint must be reformulated slightly.

For a normal distribution with non-zero mean $\bm{\mu}$, \eqref{eq:cc_formulation} must be adapted to
\begin{IEEEeqnarray}{rl}
\IEEEyesnumber
\bm{g}^\top_k \bm{z}_k &\leq h_k - \mathrm{E}\lr{\bm{g}^\top_k \bm{e}_k}  - \gamma_k \IEEEyessubnumber \\
\mathrm{Pr} \left(\bm{g}^\top_k \bm{e}_k - \mathrm{E}\lr{\bm{g}^\top_k \bm{e}_k} \leq \gamma_k \right) &= \gls{rp} \IEEEyessubnumber
\end{IEEEeqnarray}
where the scalar expectation value $\mathrm{E}\lr{\bm{g}^\top_k \bm{e}_k}$ represents an additional constraint tightening to account for the non-zero mean $\bm{\mu}$.

\appendix

\section{Probability Theory Basics}
\label{sec:preliminaries}

\subsection{Normal Distributions}
\label{sec:norm_dist}

A normal distribution (also known as a Gaussian distribution) is characterized by its mean $\bm{\mu}$ and its covariance $\bm{\Sigma}$, abbreviated by $\mathcal{N}\sim(\bm{\mu},\bm{\Sigma})$. A special case is given by the standard normal distribution where $\bm{\mu} = \bm{0}$ and $\bm{\Sigma} = \bm{I}$ with identity matrix~$\bm{I}$. In the following, we consider the univariate normal distribution with mean $\mu$, variance $\sigma^2$, and standard deviation $\sigma$.

The \gls{pdf} 
is given by
\begin{IEEEeqnarray}{c}
f(x) = \frac{1}{\sigma} \frac{1}{\sqrt{2 \pi}} \mathrm{exp} \left( - \frac{1}{2} \left( \frac{x - \mu}{\sigma} \right) \right)
\end{IEEEeqnarray}
where it holds that 
\begin{IEEEeqnarray}{c}
\int_{- \infty}^{\infty} f(x) \mathrm{d}x = 1 .
\end{IEEEeqnarray}
The factor $\frac{1}{\sigma}$ accounts for the adjusted variance compared to the standard normal distribution.

For a continuous distribution, the \gls{pdf} at a specific point $x$ does not yield a probability. Therefore, we introduce the \gls{cdf} with random variable $X$
\begin{IEEEeqnarray}{rl}
\IEEEyesnumber
\mathrm{Pr}(X \leq x) = \Phi(x) = \int_{- \infty}^{x} f(x) \mathrm{d}x &= \frac{1}{2} \left[ 1 + \mathrm{erf} \left( \frac{x- \mu}{\sigma \sqrt{2}} \right) \right] \IEEEyessubnumber \\
&= \frac{1}{2}  + \frac{1}{2} \mathrm{erf} \left( \frac{x- \mu}{\sigma \sqrt{2}} \right) . \IEEEyessubnumber
\end{IEEEeqnarray}
For $x = \mu$, the cumulative probability is exactly $0.5$.

In the \gls{cdf}, only one side of the distribution is considered. If we consider both sides, we obtain
\begin{IEEEeqnarray}{rl}
\IEEEyesnumber
\mathrm{Pr}(-x \leq X \leq x) &= \int_{-x}^{x} f(x) \mathrm{d}x \IEEEyessubnumber\\
&= \int_{-x}^{0} f(x) \mathrm{d}x + \int_{0}^{x} f(x) \mathrm{d}x \IEEEyessubnumber\\
&= 2 \int_{0}^{x} f(x) \mathrm{d}x \IEEEyessubnumber\\
&= 2 \left( \int_{-\infty}^{x} f(x) \mathrm{d}x - \int_{-\infty}^{0} f(x) \mathrm{d}x \right) \IEEEyessubnumber\\
&=  2 \left( \left(\frac{1}{2}  + \frac{1}{2} \mathrm{erf} \left( \frac{x- \mu}{\sigma \sqrt{2}} \right) \right) - \frac{1}{2}  \right) \IEEEyessubnumber\\
&= \mathrm{erf} \left( \frac{x- \mu}{\sigma \sqrt{2}} \right)  .\IEEEyessubnumber 
\end{IEEEeqnarray}

\subsection{Chebyshev's Inequality}
\label{sec:chebyshev}

Chebyshev's inequality provides a bound for probability distributions. Chebyshev's inequality indicates how likely it is that a random variable differs from the mean given a certain threshold $c$. 

For zero mean distributions, i.e., $\mu = 0$, and variance $\sigma^2$, it holds that
\begin{IEEEeqnarray}{c}
\mathrm{Pr} \left( |X| \geq c \right) \leq \frac{\sigma^2}{c^2}
\end{IEEEeqnarray}
where $c > 0$. Complementary, it holds that
\begin{IEEEeqnarray}{c}
\mathrm{Pr} \left( |X| < c \right) \geq 1 - \frac{\sigma^2}{c^2} .
\end{IEEEeqnarray}
Note the change from $\geq$ to $<$. 

The zero mean results may be extended to distributions with mean $\mu \neq 0$, yielding
\begin{IEEEeqnarray}{rl}
\IEEEyesnumber
\mathrm{Pr} \left( |X - \mu| \geq c \right) &\leq \frac{\sigma^2}{c^2} \IEEEyessubnumber \\
\mathrm{Pr} \left( |X - \mu| < c \right) &\geq 1 - \frac{\sigma^2}{c^2} . \IEEEyessubnumber
\end{IEEEeqnarray}

\paragraph{Cantelli's Inequality}

The one-sided Chabyshev's inequality is also known as the Cantelli's inequality where only a single tail of the distribution is considered. Cantelli's inequality for zero mean and non-zero mean distributions is given by
\begin{IEEEeqnarray}{rl}
\IEEEyesnumber
\mathrm{Pr} \left( X \geq c \right) &\leq \frac{\sigma^2}{\sigma^2 + c^2} \IEEEyessubnumber \\
\mathrm{Pr} \left( X - \mu \geq c \right) &\leq \frac{\sigma^2}{\sigma^2 + c^2} . \IEEEyessubnumber 
\end{IEEEeqnarray}
from which it follows that
\begin{IEEEeqnarray}{rl}
\IEEEyesnumber
\mathrm{Pr} \left( X  < c \right) = 1 - \mathrm{Pr} \left( X - \mu \geq c \right) \geq 1 - \frac{\sigma^2}{\sigma^2 + c^2} \IEEEyessubnumber \\
\mathrm{Pr} \left( X - \mu < c \right) = 1 - \mathrm{Pr} \left( X - \mu \geq c \right) \geq 1 - \frac{\sigma^2}{\sigma^2 + c^2}. \IEEEyessubnumber 
\end{IEEEeqnarray}
Note that $a \leq b 	\Longleftrightarrow -a \geq -b$.

\newpage
\bibliography{Dissertation_bib}
\bibliographystyle{alpha}

\end{document}